# OVERVIEW OF BEAM DYNAMICS STUDIES AT DAΦNE

M. Zobov, INFN Laboratori Nazionali di Frascati, Frascati, Italy
for DAΦNE Collaboration Team*

*Abstract*

Since several years the DAΦNE Team has been discussing ideas and performing experimental activities aimed at the collider luminosity increase. In this paper we briefly describe the proposed ideas and discuss results of the most relevant beam dynamics experimental studies that have been carried at DAΦNE. We also introduce the concept of crab waist collisions that is the base of the undergoing DAΦNE upgrade.

## INTRODUCTION

DAΦNE is an electron-positron collider working at the c.m. energy of the Φ resonance (1.02 GeV) to produce a high rate of K mesons [1]. The collider complex consists of two independent rings having two common Interaction Regions (IR) and an injection system composed of a full energy linear accelerator, a damping/accumulator ring and transfer lines. Figure 1 shows a view of the DAΦNE accelerator complex while some of the main collider parameters are listed in Table 1.

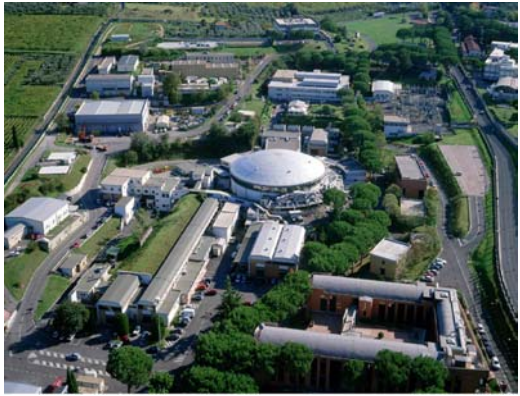

Fig. 1 View of DAΦNE accelerator complex.

Since 2000 DAΦNE has been delivering luminosity to three experiments, KLOE [2], FINUDA [3] and DEAR [4]. The KLOE experimental detector surrounded by a superconducting solenoid has been used for a wide variety of physics measurements with emphasis on the kaon decays, and most notably on the issue of CP violation. The second magnetic detector FINUDA is devoted to the study of hypernuclear physics. The small non-magnetic experiment DEAR has been used for the study of the properties of kaonic atoms.

In 2007 DAΦNE was shut down for the SIDDHARTA experiment installation [5] and for relevant collider modifications aimed at testing the novel idea of crab waist collisions [6, 7, 8]. DAΦNE operations with the crab waist scheme started in the very end of 2007 and the first results of the new scheme implementation are reported in [9] at this Workshop.

Table 1: DAΦNE main parameters (KLOE run)

| | |
|---|---|
| Energy [GeV] | 0.51 |
| Trajectory length [m] | 97.69 |
| RF frequency [MHz] | 368.26 |
| Harmonic number | 120 |
| Damping time, $\tau_E/\tau_x$ [ms] | 17.8/36.0 |
| Bunch length [cm] | 1-3 |
| Number of colliding bunches | 111 |
| Beta functions $\beta_x/\beta_y$ [m] | 1.6/0.017 |
| Emittance, $\varepsilon_x$ [mm·mrad] (KLOE) | 0.34 |
| Coupling [%] | 0.2-0.3 |
| Max. tune shifts | 0.03/0.04 |
| Max. beam current e-/e+ [A] | 2.5/1.4 |

## COLLIDER PERFORMANCE

Since the beginning of the experimental data taking runs in 2000 DAΦNE has been continuosly improving its performances in terms of luminosity, lifetime and backgrounds. Fig.2 shows the daily peak luminosity for KLOE, DEAR and FINUDA runs.

The DEAR experiment was performed in less than 5 months in 2002-2003, collecting about 100 pb$^{-1}$, with a peak luminosity of $0.7 \times 10^{32}$ cm$^{-2}$s$^{-1}$. The KLOE experimental program has been completed in 2006, acquiring more than 2 fb$^{-1}$ on the peak of the Φ resonance, more than 0.25 fb$^{-1}$ off-resonance and performing a high statistics resonance scan. The best peak luminosity obtained during this run was $1.5 \times 10^{32}$ cm$^{-2}$s$^{-1}$, with a maximum daily integrated luminosity of about 10 pb$^{-1}$. The second run of FINUDA, which collected 0.96 fb$^{-1}$, started in October 2006. During this run a peak luminosity of $1.6 \times 10^{32}$ cm$^{-2}$s$^{-1}$ has been achieved, while a maximum daily integrated luminosity similar to that in the KLOE run has been obtained with lower beam currents, lower number of bunches and higher beta functions at the collision point.

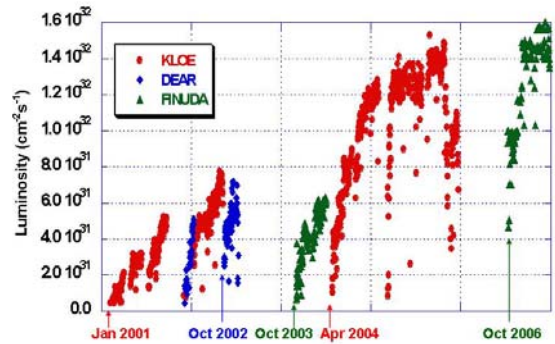

Figure 2. DAΦNE peak luminosity for KLOE (red), DEAR (blue) and FINUDA (green).

The steady luminosity progress shown in Fig. 2 was achieved by the optimization of the machine parameters and hardware changes implemented during long shut downs in 2003 and in 2006. In the following we overview only some of the main beam dynamics issues that have been studied at DAΦNE and helped to improve the collider performance.

## SINGLE BUNCH DYNAMICS

We have not detected any destructive single bunch instability. Indeed, we have managed to store bunch currents that are by a factor 10 higher than those required in typical operating conditions. So the main efforts in single bunch dynamics optimization were aimed at bunch lengthening reduction and single bunch vertical size blow up elimination.

Nowdays analytical estimates (see, for example, [10, 11]) and existing 2D and 3D numerical codes have proven their reliability in RF design and vacuum chamber's coupling impedance optimization. Careful vacuum chamber design and detailed impedance calculations allow to predict bunch behaviour in modern storage rings [12]. A good agreement between bunch lengthening simulations and measurements was found for all the three rings of the DAΦNE accelerator complex, the accumulator [13] and the electron and positron main rings [14, 15]. In particular, for the main rings it was possible not only to calculate correctly the bunch lengthening effect and the bunch charge distribution, but also to predict in advance the bunch shape oscillation instability. This quadrupole mode instability was clearly observed in experimental measurements [16].

Achieving higher luminosity required shorter bunch length, i.e. a reduction of the beam coupling impedance responsible for the bunch lengthening in the main rings. This was particularly important for the electron ring where the impedance was by about a factor 2 higher due to the ion clearing electrodes (see Fig. 3).

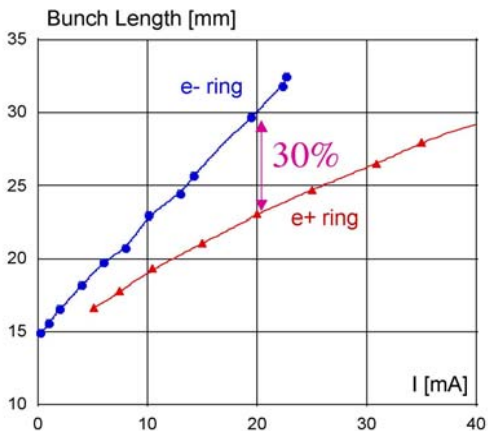

Fig. 3 Bunch length in the electron (blue) and positron (red) rings as a function of bunch current (measured at $V_{RF}$=120 kV, $\alpha_c$ = 0.017).

Another harmful effect related to the impedance was the vertical beam size blow up. As can be seen in Fig. 4, the blow up was stronger for higher RF voltages and higher bunch currents and it had a threshold behavior. Dedicated measurements have shown that the blow up threshold was strongly correlated with the longitudinal microwave instability threshold.

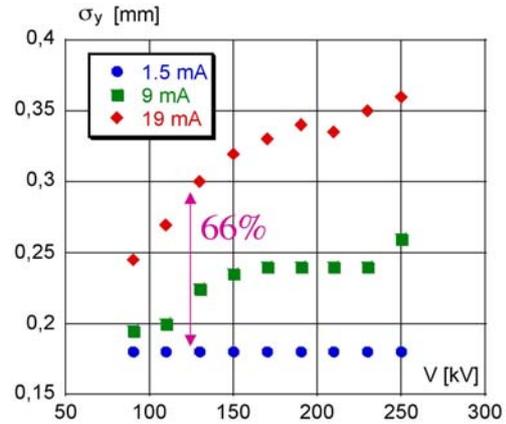

Fig. 4 Vertical beam size measured by the synchrotron light monitor as a function of RF voltage for three different bunch currents.

For the reasons above, it has been decided to decrease the electron ring coupling impedance by extracting the long ion clearing electrodes from the wiggler sections [17]. The electrodes removal is estimated to decrease the impedance by a factor 2. Its effects are:

- electron bunches are by about 25-30% shorter in typical operating conditions (see Fig. 5);
- the quadrupole instability threshold has been pushed beyond the operating bunch currents;
- no vertical beam size blow up has been observed for the whole range of operating bunch currents and RF voltages.

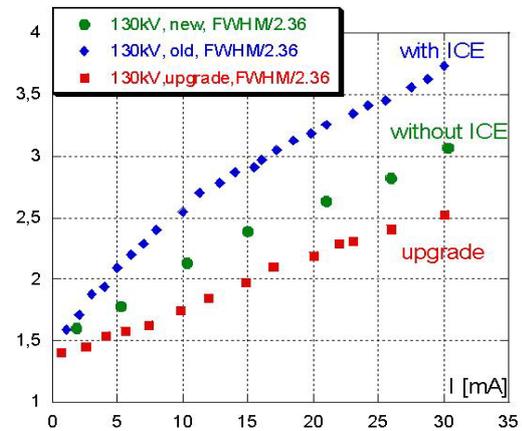

Fig. 5 Bunch lengthening (cm) in the DAΦNE electron ring at $V_{RF}$ = 130 kV, $\alpha_c$ = 0.019 (blue – original vacuum chamber, green – after clearing electrode removal, red – upgraded vacuum chamber)

This helped to increase the specific luminosity by approximately 50% during the last FINUDA run [18]. During the upgrade for the crab waist experiment the DAΦNE vacuum chamber was substantially modified (see [19]). In particular, the design of both interaction regions was greatly simplified. Now the IR chambers are just two straight intersecting tubes with very smooth tapers. Besides, new low impedance injection kickers and new shielded bellows were installed in the rings. These modifications have reduced further the beam coupling impedance and decreased bunch lengthening, as one can see in Fig. 5.

## MULTIBUNCH DYNAMICS

Since the very beginning of the DAΦNE project multibunch instabilities were recognized as one of the most harmful potential dangers. For this reason the vacuum chamber was designed taking care of HOM damping and shifting the HOM frequencies far from revolution harmonics in order to avoid excessive power losses. New designs and novel ideas were adopted for almost all principal vacuum chamber components: RF cavities [20, 21], shielded bellows [22], longitudinal feedback kickers [23], BPMs [24], DC current monitors, injection kickers, transverse feedback kickers and others [25]. For example, longitudinal feedback kickers based on the DAΦNE kicker design are routinely used in more than 10 operating colliders and synchrotron radiation sources.

However, these measures were not sufficient to eliminate completely any multibunch instability and powerful feedback systems also were developed to damp the residual unstable multibunch oscillations.

During DAΦNE operation two multibunch instabilities were limiting collider performance: the longitudinal quadrupole instability in the electron ring and the strong horizontal instability in the positron one.

The multibunch longitudinal quadrupole instability was essentially the same as the single bunch quadrupole mode, but much stronger in multibunch operation. For some time this kind of instability was limiting the electron beam current at a level of 700-800 mA due to the current saturation at injection and was leading to luminosity loss in beam-beam collisions of unstable bunches. The instability was kept under control by a proper feedback tuning providing different longitudinal kicks for the bunch heads and bunch tails [16]. Later, due to the impedance reduction after ion clearing electrodes removal the instability has been completely suppressed for the whole range of operating beam currents.

Instead the positron beam horizontal instability still remains one of the most stringent collider performance limitations. Indeed, the maximum positron stored current is significantly lower than the electron one. The instability has many typical features of electron cloud instability:
- Large positive tune shift
- Anomalous pressure rise
- Very fast growth rate (several μs)
- Bunch pattern dependence (see Fig. 6)
- Some evidence of beam scrubbing
- Others

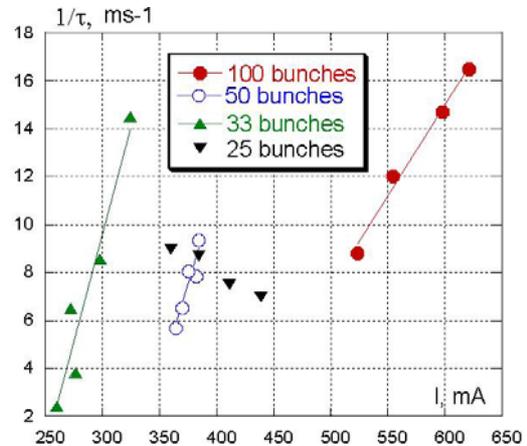

Fig. 6. Positron beam instability decrements as a function of beam current for different bunch patterns.

Several measures helped to keep the instability under control to store reasonable positron beam currents in collision (>1 A). These are:
- Implementation of powerful transverse feedback systems providing damping times as fast as few μsec [26].
- Careful injection optimization necessary to minimize perturbation of the already stored beam. It has been achieved by reducing the injection kicker pulse length and by fine injection bump adjustment.
- During the current DAΦNE run for SIDDHARTA experiment several solenoids have been wound around the vacuum chambers of interaction regions and straight sections. This yielded some reduction of anomalous pressure rise and has also been beneficial to the increase of the vertical instability threshold.
- A modest effect of beam scrubbing has been observed on a long time scale.

Further improvements are expected after installation of additional solenoids. Besides, the feasibility of vacuum chamber titanium coating is under discussion.

## BEAM-BEAM COLLISIONS

For lepton colliders with low transverse impedance the existing beam-beam simulation codes predict rather well beam-beam behaviour of colliding bunches: beam blow up, beam-beam lifetime, good working point areas etc. In order to simulate beam-beam effects in DAΦNE the LIFETRAC [27] code has been successfully used.

Fig. 7 shows the luminosity tune scan plot provided by LIFETRAC where the brighter red colours correspond to higher luminosity. The superimposed solid line arrow shows how the DAΦNE tunes changed with time, while the stars indicate the electron and positron ring tune during the FINUDA run. Following the code predictions, optimizing dynamic aperture and going closer to the integer tunes we have gradually improved both the peak luminosity and beam-beam lifetime.

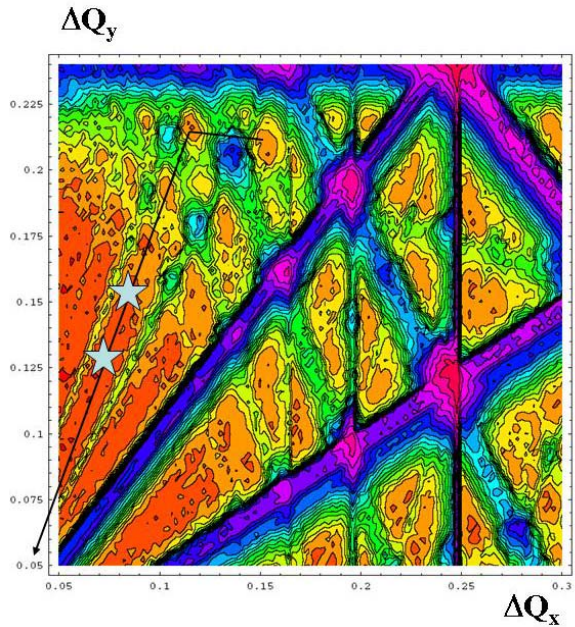

Fig. 7 Luminosity numerical tune scan.

At an early stage of beam-beam studies it was predicted that the crosstalk between lattice nonlinearities and beam-beam interaction could dramatically deteriorate the collider performance [28]. This effect becomes particularly strong in multibunch operation. It has been shown numerically that in order to cope with this effect it is necessary to reduce the nonlinear tune dependence on oscillation amplitude as much as possible.

For this purpose three octupoles have been installed in each ring. They have proven to be very effective in increasing the dynamic aperture and in compensating the cubic nonlinearity of the lattice. The octupoles have become essential in multibunch collisions of consecutive bunches to prevent strong lifetime reduction at the maximum beam currents [29].

Long-range beam-beam interactions (parasitic crossings) were another source of luminosity degradation in the DAΦNE original configuration before the collider upgrade for crab waist collisions. Due to drastic lifetime reduction the parasitic crossings put a limit on the maximum storable currents and, as a consequence, on the achievable peak and integrated luminosity. In order to mitigate this problem, numerical and experimental studies of the parasitic crossings compensation by current-carrying wires have been done [29, 30]. During the operation for the KLOE experiment two such wires have been installed at both ends of the interaction region, outside the vacuum chamber. They produced a relevant improvement in the lifetime of the "weak" beam (positrons) at the maximum current of the "strong" one (electron) without luminosity loss (see Fig. 8), in agreement with the numerical predictions.

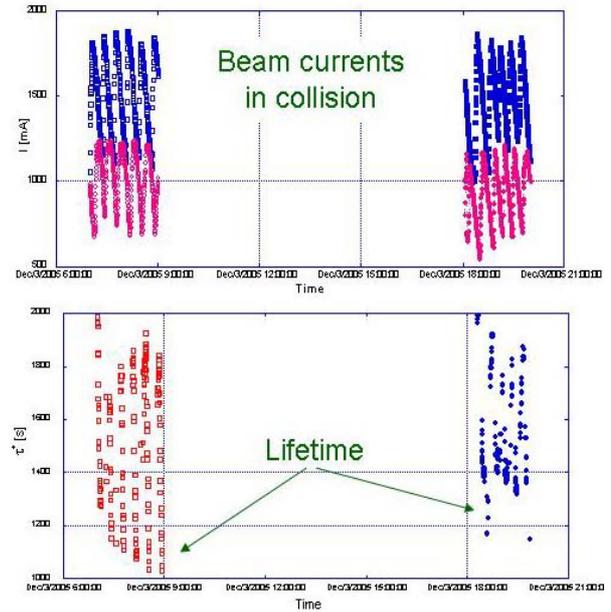

fig. 8 Beam currents and positron beam lifetime with current-carrying wire off (left) and on (right).

## DAΦNE UPGRADE ACTIVITIES

Since several years the DAΦNE team has been discussing proposals and ideas aimed at increasing the luminosity of the collider. Some of them are listed below:
- "All wiggler" structure to decrease the damping time
- Negative momentum compaction factor [31]
- PC compensation with current-carrying wires [30]
- Collisions under a very high crossing angle [32]
- Strong RF focusing [33, 34]
- Crabbed waist collisions [7, 35].

Collisions in a negative momentum compaction lattice and PC compensation with the wires have been tested experimentally at DAΦNE while 3 different proposals for DAΦNE upgrade, DANAE [36], strong RF focusing collider [37] and crab waist upgrade [35], have been published.

After many discussions the DAΦNE team decided, and the LNF Scientific Committee approved, testing the large crossing angle and crab waist compensation scheme at DAΦNE since:
- A high luminosity gain is expected
- Does not require large hardware modifications
- Fits the DAΦNE schedule: the modifications are being done during the planned shut down for the FINUDA detector roll-out and SIDDHARTA detector installation
- The cost of the modifications is moderate
- If the test is successful the crabbed waist scheme can be adopted for the SuperB factory design [38].

The Φ-factory has been upgraded in the second half of 2007 to implement the crab waist collision scheme [19]. Commissioning of the modified collider started in November 2007. Details of the DAΦNE hardware upgrade and the first obtained results are described elsewhere at this Workshop [9]. Below we discuss the crab waist collision concept from beam dynamics point of view.

*Crab Waist Concept*

The Crab Waist scheme of beam-beam collisions can substantially increase collider luminosity since it combines several potentially advantageous ideas. Let us consider two bunches with the vertical $\sigma_y$, horizontal $\sigma_x$ and longitudinal $\sigma_z$ sizes colliding under a horizontal crossing angle $\theta$ (as shown in Fig. 9a). Then, the CW principle can be explained, somewhat artificially, in the three basic steps.

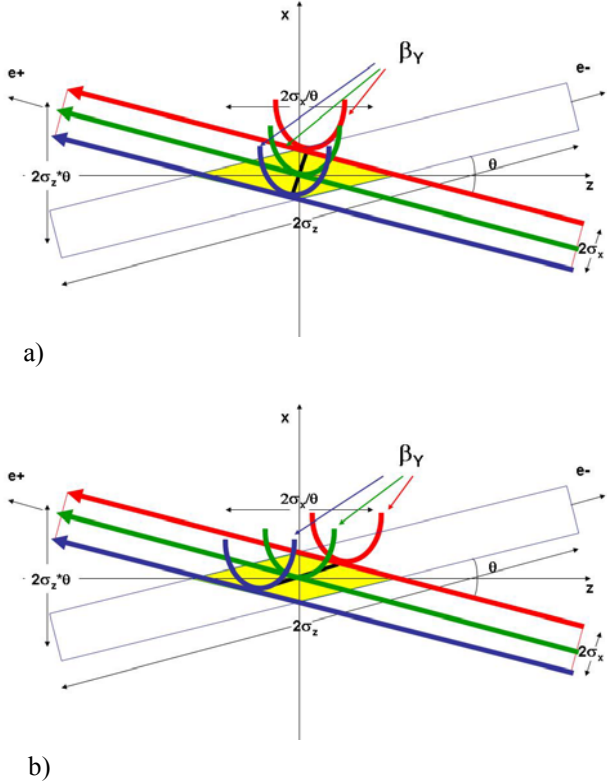

a)

b)

Fig. 9 Crab Waist collision scheme
((a) – crab sextupoles off; (b) – crab sextupoles on)

The first one is large Piwinski angle. For collisions under a crossing angle $\theta$ the luminosity $L$ and the horizontal and vertical tune shifts $\xi_x$ and $\xi_y$ scale as (see, for example, [39]):

$$L \propto \frac{N\xi_y}{\beta_y^*} \propto \frac{1}{\sqrt{\beta_y^*}}; \quad \xi_y \propto \frac{N\sqrt{\beta_y^*}}{\sigma_z \theta}; \quad \xi_x \propto \frac{N}{(\sigma_z \theta)^2}$$

Here the Piwinski angle is defined as:

$$\phi = \frac{\sigma_z}{\sigma_x} tg\left(\frac{\theta}{2}\right) \approx \frac{\sigma_z}{\sigma_x}\frac{\theta}{2}$$

with $N$ being the number of particles per bunch. Here we consider the case of flat beams, small horizontal crossing angle $\theta \ll 1$ and large Piwinski angle $\phi \gg 1$.

The idea of colliding with a large Piwinski angle is not new (see, for example, [40]). It has been also proposed for hadron colliders [41, 42] to increase the bunch length and the crossing angle. In such a case, if it were possible to increase $N$ proportionally to $\sigma_z\theta$, the vertical tune shift $\xi_y$ would remain constant, while the luminosity would grow proportionally to $\sigma_z\theta$. Moreover, the horizontal tune shift $\xi_x$ drops like $1/\sigma_z\theta$. However, differently from [41, 42], in the crab waist scheme described here the Piwinski angle is increased by decreasing the horizontal beam size and increasing the crossing angle. In this way we can gain in luminosity as well, and the horizontal tune shift decreases due the larger crossing angle. But the most important effect is that the overlap area of the colliding bunches is reduced, since it is proportional to $\sigma_x/\theta$ (see Fig. 9).

Then, as the second step, the vertical beta function $\beta_y$ can be made comparable to the overlap region length (i.e. much smaller than the bunch length):

$$\beta_y^* \approx \frac{\sigma_x}{\theta} \ll \sigma_z$$

We get several advantages in this case:
- Small spot size at the IP, i.e. higher luminosity L.
- Reduction of the vertical tune shift $\xi_y$.
- Suppression of synchrobetatron resonances [43].
- Reduction of the vertical tune shift with the synchrotron oscillation amplitude [43].

There are additional advantages in such a collision scheme: there is no need to decrease the bunch length to increase the luminosity as proposed in standard upgrade plans for B- and Φ-factories [44, 45, and 36]. This will certainly helps solving the problems of HOM heating, coherent synchrotron radiation of short bunches, excessive power consumption etc. Moreover, parasitic collisions (PC) become negligible since with higher crossing angle and smaller horizontal beam size the beam separation at the PC is large in terms of $\sigma_x$.

However, large Piwinski angle itself introduces new beam-beam resonances which may strongly limit the maximum achievable tune shifts (see [46], for example). At this point the crab waist transformation enters the game boosting the luminosity. This is the third step. The transformation is described by the Hamiltonian

$$H = H_0 + \frac{1}{2\theta} x p_y^2$$

Here $H_0$ is the Hamiltonian describing particle's motion without CW; $x$ the horizontal coordinate, $p_y$ the vertical momentum. Such a transformation produces a rotation of the vertical beta function according to:

$$\beta_y = \beta_y^* + \frac{(s - x/\theta)^2}{\beta_y^*}$$

As shown in Fig. 9b, in this case the beta function waist of one beam is oriented along the central trajectory of the other one.

The crab waist transformation yields a small geometric luminosity gain due to the vertical beta function redistribution along the overlap region. It is estimated to be of the order of several percent [47]. However, the dominating effect comes from the suppression of betatron (and synchrobetatron) resonances arising (in collisions without CW) through the vertical motion modulation by the horizontal oscillations [7, 8]. In practice the CW vertical beta function rotation is provided by sextupole magnets placed on both sides of the IP in phase with the IP in the horizontal plane and at π/2 in the vertical one (as shown in Fig. 10).

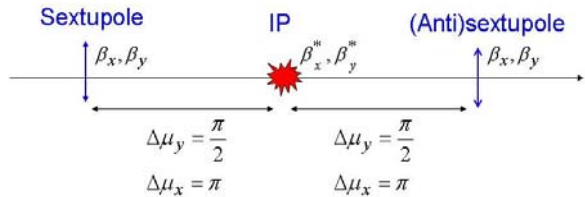

Fig. 10. Crab sextupole locations.

The crab sextupole strength should satisfy the following condition depending on the crossing angle, the beta functions at the IP and the sextupole locations:

$$K = \frac{1}{2\theta} \frac{1}{\beta_y^* \beta_y} \sqrt{\frac{\beta_x^*}{\beta_x}}$$

A detailed discussion on the resonance suppression in crab waist collisions with many numerical examples can be found in [48], while a semi-analytical treatment of this effect is given in [49, 50].

## CONCLUSIONS

Since 2000 DAΦNE has been delivering luminosity to three physics experiments, KLOE, FINUDA and DEAR, steadily improving luminosity, lifetime and backgrounds. Beam dynamics studies have been one of the key contributions to the successful collider operation.

During last years the DAΦNE Team has been discussing proposals and ideas for the Φ-factory luminosity increase. Some of them, such as beam-beam collisions with negative momentum compaction factor, parasitic crossings compensation with current-carrying wires and others have been tested experimentally.

After many discussions and thorough analytical and numerical beam dynamics studies it has been decided to implement a novel scheme of crab waist collisions, exploiting the shut down for the SIDDHARTA experiment installation. With this new collision configuration it is expected to reach a luminosity of the order of $5 \times 10^{32}$ cm$^{-2}$s$^{-1}$.

[*]D.Alesini, D.Babusci, S.Bettoni, M.Biagini, C.Biscari, R.Boni, M.Boscolo, F.Bossi, B.Buonomo, A.Clozza, G.Delle Monache, T.Demma, G.Di Pirro, A.Drago, A.Gallo, A.Ghigo, S.Guiducci, C.Ligi, F.Marcellini, C.Marchetti, G.Mazzitelli, C.Milardi, F.Murtas, L.Pellegrino, M.Preger, L.Quintieri, P.Raimondi, R.Ricci, U.Rotundo, C.Sanelli, M.Serio, F.Sgamma, B.Spataro, A.Stecchi, A.Stella, S.Tomassini, C.Vaccarezza, M.Zobov (LNF, Italy); I.Koop, E.Levichev, S.Nikitin, P.Piminov, D.Shatilov, V.Smaluk (BINP, Russia); J.Fox, D.Teytelman (SLAC, USA); K. Ohmi (KEK, Japan)[*]